\documentclass{aa}
\usepackage{amssymb,amsfonts,amsmath,times,pict2e}
\usepackage{natbib}
\usepackage{graphicx}
\usepackage{enumerate}
\usepackage{subfigure}
\usepackage{bm}
\usepackage{float}

\usepackage{color}
\usepackage{ulem}
\definecolor{mygray}{gray}{0.6}
\definecolor{magenta}{rgb}{0.858, 0.188, 0.478}

\newcommand{\Rmnum}[1]{\expandafter\@slowromancap\romannumeral #1}

\usepackage{xspace}
\newcommand{\fg}[1]{Fig.~\ref{fig:#1}}
\newcommand{\eq}[1]{Eq.~(\ref{eq:#1})}
\newcommand{\se}[1]{Sect.~\ref{sec:#1}}
\newcommand{\tb}[1]{Table~\ref{tab:#1}}

\newcommand{\AU}{ \  \rm AU}

\newcommand{\Ms}{ \  \rm M_\odot }
\newcommand{\Msyr}{ \  \rm M_\odot yr^{-1} }
\newcommand{\yr}{ \   \rm yr}

\newcommand{\kG}{ \ \rm kG}

\newcommand{\taud}{ \tau_{\rm d}}

\newcommand{\mdot}{ \dot M_{\rm g0}}

\begin{document}

\title{Dynamical rearrangement of super-Earths during disk dispersal}
\subtitle{II. Assessment of the magnetospheric rebound model for planet formation scenarios }

\author{ Beibei Liu \inst{1}, Chris W. Ormel\inst{1} }
\institute{Anton Pannekoek Institute (API), University of Amsterdam, Science Park 904,1090GE Amsterdam, The Netherlands\label{inst1}\\
\email{b.liu@uva.nl,c.w.ormel@uva.nl}
 }

\date{\today}

\abstract{
The Kepler mission has provided a large sample to statistically analyze the orbital properties of the super-Earth planet population. We hypothesize that these planets formed early and consider the problem of matching planet formation theory to the current orbital configurations. Two scenarios -- disk migration and \textit{in-situ} formation -- have been proposed to explain the origin of these planets. In the migration scenario, planets migrate inward to the inner disk  due to  planet-disk interaction, whereas in the \textit{in-situ} scenario planets assemble locally. Therefore,  planets formed by migration are expected to end up in resonances, whereas those formed \textit{in-situ} are expected to stay in short period ratios and in non-resonant orbits. Both predictions are at odds with observations.
}
%aims
{  We investigate whether a preferred formation scenario can be identified through a comparison between the magnetospheric rebound model and the Kepler data.}
% methods
{We conduct N-body simulations of two-planet systems during the disk dispersal phase. Several distributions of model parameters are considered and we make a statistical comparison between the simulations and the Kepler observations.}
% results
{ 
Comparing the  migration and the \textit{in-situ} scenarios, we find that magnetospheric rebound tends to erase the difference in the orbital configuration that was initially presented.
After disk dispersal, not all planets are in resonance in the migration scenario, whereas planets do not remain in compact configurations in the \textit{in-situ} scenario. In both scenarios, the orbits of planets increase with the cavity expansion, and their period ratios have a wider distribution.
}
 %conclusion
{From a statistical perspective, the magnetospheric rebound model reproduces several observed properties of Kepler planets,  such as the fact that a significant number of planets are not in resonances and  planet pairs can end up at large period ratios. The disparity in orbital configuration between the two formation scenarios is substantially reduced after  disk dispersal.  } 

\keywords{methods: numerical – planets and satellites: formation – planet–disk interactions}

\maketitle

\section{Introduction} 
\label{sec:intro}
The Kepler and K2 missions have vastly increased the number of  detected planets and revolutionized our understanding of planet formation. One of the most intriguing findings is that super-Earths (defined here as radii $R_{\rm p} \leq 4 R_{\oplus}$ with orbital periods $P \lesssim100$ days) are the most abundant type of planets.   
Since the Kepler mission has discovered so many super-Earth planets, the dynamical configuration of these planets can be analyzed from a statistical perspective.

We summarize here the key properties of super-Earths (see \cite{Winn2015} for a review). First, super Earths are very common. The occurrence of super-Earths among solar-type stars is around $50\%$  \citep{Petigura2013} with a slight dependence on stellar type \citep{Howard2012,Fressin2013,Mulders2015}, but independent of stellar metallicity \citep{Wang2013, Buchhave2014}. These findings are also consistent with  ground-based radial velocity  surveys \citep{Mayor2011,Bonfils2013}.  Second, super-Earths are commonly discovered in multiple, compact, and short period systems with relatively low eccentricities and inclinations \citep{Fang2012,Johansen2012,Fabrycky2014,VanEylen2015,Shabram2016,Xie2016}.  Third, the distribution of their period ratios is neither  strongly peaked at resonances, nor entirely uniformly spaced \citep{Fabrycky2012}.    
Fourth, some super-Earth are inferred to contain H/He-rich envelopes. We can deduce the composition of a few dozen Kepler super-Earth planets from transit-timing variation \citep{Lithwick2012b} or RV follow-up surveys \citep{Marcy2014}. The bulk densities of many of these planets are too low to be consistent with a purely rocky (Earth-like) composition. Detailed modeling of mass-radius relationships indicates that these super-Earths contain non-negligible H/He envelopes \citep{Lopez2014}. They may have already formed in the early gas-rich disk phase, and also do not suffer violent giant-impacts after the depletion of disk gas \citep{Inamdar2015}.

For the solar system, local formation models  (see \cite{Raymond2014b,Morbidelli2016} for reviews and references therein) attempt to explain the origin of terrestrial planets. These models are formulated according to the minimum-mass solar nebula (MMSN) profile for the surface density \citep{Weidenschilling1977b,Hayashi1981}. Extending the local formation paradigm to extrasolar systems,  the minimum-mass extrasolar nebula (MMEN) disk model has been constructed  \citep{Chiang2013}. However, $\Sigma_\mathrm{MMEN}$ is much larger than $\Sigma_\mathrm{MMSN}$, approximately by a factor $10$.  Apart from the large disk mass, the surface density profiles of MMENs are inconsistent with  sub-millimeter observations of the outer disks around  young T Tauri stars \citep{Andrews2009,Raymond2014}.
In addition, \cite{Schlichting2014} demonstrated that for many of these systems the gas disks are likely to be gravitationally unstable, assuming entirely local formation and a standard gas-to-dust ratio.  She concluded that planets or their building blocks at further distances were transported to the inner disk.

 Here we consider two scenarios that could explain the origin of super-Earth planets. The disk migration scenario \citep{Terquem2007,McNeil2010,Cossou2014,Coleman2014,Coleman2016,Ogihara2015,Liu2015,Liu2016,Izidoro2017,Ormel2017} assumes that (proto)planets have already formed in the outer part of the disk. They then undergo inward Type I migration (see \cite{Kley2012} for a review) until reach the inner disk edge, where the stellar magnetic field truncates the disk \citep{Lin1996}. On the other hand, the \textit{in-situ} scenario assumes that super-Earths form in the inner part of the disk without (substantial) disk migration \citep{Hansen2012,Hansen2013,Lee2014,Dawson2016}. Importantly, we would like to emphasize that our definition of \textit{in-situ} formation allows for material to be brought in from the outer disk, in the form of pebbles  \citep{Weidenschilling1977a} or planetesimals \citep{Grishin2015}. These drifting pebbles stop and accumulate at pressure bumps in the inner disk region \citep{Chatterjee2014,Hu2016}.  Such a large concentration of solids can therefore trigger streaming instability to generate planetesimals \citep{Youdin2005,Johansen2007}.  Subsequent cohesive collisions among planetesimals yield the super-Earth planets.    
One of the goals in this paper is to investigate whether we can discriminate between the \textit{in situ} and migration scenarios (also see \cite{Ogihara2015,Lee2017}).

In \cite{Liu2017} (referred to as Paper I hereafter), we proposed a new mechanism -- magnetospheric rebound --  which can rearrange the dynamical configuration of super-Earths during the gas disk dispersal phase. We constructed a disk model with an inner magnetospheric cavity, which expands during disk dispersal. The Type I torque that the planet experiences is negative when the planet is far away from the cavity radius. But the torque becomes positive near the cavity radius, which stalls the inward migration. Adopting analytical expressions of  the Type I torque, we conducted N-body simulations on the evolution of  two-planet systems. As the disk gas depletes, the planets migrate outward with the cavity expansion. When planets decouple from the cavity radius, they end up in a non-resonant configuration.    As one of our key results, we found that our model predicts a correlation between the period ratio of two planets and their mass ratio.   
  
In this work (Paper II) we will  study the consequences of the magnetospheric rebound model for the above two super-Earth formation scenarios. The paper is structured as follows. 
In \se{model}, we briefly review our model in Paper I, and then outline the initial conditions of the two formation scenarios. For both scenarios we statistically simulate the evolution of the Kepler two-planet systems during the gas-disk depletion (\se{parameter}). In \se{optimization}, we compare our model results with the observations by investigating different distributions of input parameters.  We summarize our findings in \se{conclusion}.

\section{Model}
\label{sec:model}

\subsection{Model description}
\label{sec:description}
We briefly discuss the magnetospheric rebound model, which is described in detail in Paper I.
The magnetospheric cavity truncates the disk at a radius,
\begin{equation}
 r_{\rm c} = 0.1  \left(  \frac{ \dot M_{\rm g}}{10^{-9}   \Ms  \yr^{-1}}\right) ^{-2/7} \left(  \frac{ B_{\ast} }{ 1  \kG}\right)^{4/7 }  \AU,
 \label{eq:r_cavity}
\end{equation}
where $ \dot M_{\rm g}$ is the gas disk accretion rate and $B_{\ast}$ is the stellar magnetic field strength at the stellar surface.
The disk is assumed to be  in a steady state such that the relation between the gas surface density $\Sigma$ and $\dot{M}_g$ reads
\begin{equation}
{\dot M}_{\rm g} =  3 \pi \Sigma \alpha_{\nu},
 \label{eq:steady}
\end{equation}
where $\alpha_{\nu}$ is the dimensionless viscous parameter \citep{Shakura1973}. 
We assume that the disk remains accreting at a constant  rate during the gas-rich phase, but that  ${\dot M}_{\rm g} $ declines exponentially with a depletion timescale ($\taud$) after  a time $\tau_0$:
 \begin{equation}
{\dot M}_{\rm g} =\begin{cases}
 {\dot M}_{\rm g0}  & \mbox{ when $ t \leq \tau_0$ };  \\
 {\dot M}_{\rm g0}   \exp \left[- (t-\tau_0)/\taud \right] & \mbox{ when $ t > \tau_0$ }. \\
 \end{cases}
\label{eq:taudep}
\end{equation} 
Here, $\mdot$ refers to the initial disk accretion rate at the onset of disk dispersal.

The model contains three disk and stellar parameters: 
 the initial gas disk accretion ($ \dot M_{\rm g0} $), the disk dispersal timescale ($\taud$), and the stellar magnetic filed strength ($B_{\ast}$).  
 Following Paper I, we also assume $\alpha_{\nu} =10^{-2}$.  We note, however, that the model results are not restricted to the specific choice for $\alpha_{\nu}$.   For instance, scaling by an arbitrary  factor $F$, a new set of parameters can be formed as  $\tilde{\alpha}_{\nu}  = F \alpha_{\nu}$, $\tilde{\dot M}_{g0} = F \dot M_{\rm g0} $ and $ \tilde{B}_{\ast} = \sqrt{F} B_{\ast} $. It follows from Eqs. (1) and (2) that the results with this new set of parameters ($\tilde{\dot M}_{\rm g}$, $\tilde{\alpha}_{\nu}$, $\tilde{B}_{\ast}$)  would be identical to those of ($\dot M_{\rm g}$, $\alpha_{\nu}$, $B_{\ast}$). 

\subsection{Formation scenarios of super-Earth planets: Migration vs. \textit{In-situ}}
\label{sec:scenarios}
We investigate two formation scenarios for super-Earth planets: the migration scenario and the \textit{in situ} scenario. These scenarios determine the initial periods (ratios) of the planets before disk dispersal. In Sects. 3 and 4,  we then show how magnetospheric rebound operates for these two scenarios. 

\begin{enumerate}[1.]
    \item In the migration scenario super-Earths form in the outer part of the disk (e.g., at or beyond the snowline). After acquiring their masses, planets undergo Type I migration toward  the inner edge of the disk. For simplicity, we adopt the local isothermal approximation for the Type I torque, and the migration is always inward (see Paper I). We also assume that the differential formation timescale of  neighboring planets is longer than their migration timescale.  Therefore, the inner planet has already migrated to the inner disk edge ($r_c$) before the outer planet starts migration. While migrating, the outer planet is able to cross  higher-order resonances easily (e.g., $3$:$1$ mean motion resonance (MMR), \cite{Quillen2006}). The $2$:$1$ MMR is the first strong resonant barrier in which planets may get trapped.  In the N-body simulation, we therefore initialize the inner planet at the inner disk edge and the outer planet at a period ratio of $2.1$, just slightly out of their $2$:$1$ MMR. 

    \item In the \textit{in-situ} scenario super-Earths accrete solid materials in the inner disk region.  These planetary building blocks could  be planetesimals  or pebbles that drifted from the outer part of the disk.  Because the differential formation timescale of these low-mass planets is shorter compared to their migration timescale,  we do not expect that embryos would undergo sufficient migration during rapid planetesimal accretion \citep{Chiang2013} or pebble accretion  \citep{Ormel2010,Lambrechts2012}. In the \textit{in situ} case, the inner planet is initially assumed to be located at the disk edge, whereas the outer planet is  separated randomly  from six to ten mutual Hill radii, as a consequence of orbital repulsion \citep{Kokubo1998}.
\end{enumerate}

\section{Fiducial case}
\label{sec:parameter}

We perform N-body simulations of two-planet systems. Apart from the stellar and planet gravitational forces, the planet-disk interaction is taken into account by adding Type I torques (see Paper I for details).
We aim to investigate the evolution of the planets during the disk dispersal phase.
While Paper I focused on mapping the architecture of individual Kepler systems, here we conduct a statistical comparison between the model outcomes and the observed Kepler population for the two formation scenarios (\se{scenarios}). Specifically, we assume that the key model parameters ($\mdot$, $\taud$ and $B_{\star}$) are each log-uniformly (or uniformly) distributed between an upper and a lower limit. We also statistically clarify the influence of the model parameters.
 
 \subsection{Fiducial case set-up} 
 \label{sec:setup}

The planetary systems are adopted from  the Kepler `Q1-Q17 DR24'  \citep{Coughlin2016}  data. \footnote[1]{ http://exoplanetarchive.ipac.caltech.edu/index.html.} Here we select the sample based on the following criteria: (1) both confirmed planets and candidates with $R_{p} < 4 \ R_{\oplus}$; (2) only two-planet systems; and (3) solar-type host stars ($0.7M_{\odot} < M_{\star} < 1.2M_{\odot}$). This results in a sample of  $318$ planetary systems. 
These criteria are motivated by the main goal of our paper: to investigate whether the Kepler data can distinguish a preferred formation scenario in light of  the magnetospheric rebound mechanism (see \se{optimization}).  The mechanism requires a non-gap opening condition, which is satisfied for low-mass planets around solar-type stars (see Section 5 of Paper I). So we restrict our sample by criteria (1) and (3). In this work, we focus on two-planet systems (criterion (2)) and their evolutions are only simulated during the gas disk dispersal phase. A two-planet system is always stable in a gas-free system when the planet's separation is larger than $2\sqrt{3}R_{\rm H}$ (Hill stability; \citealt{Gladman1993}).  Our results show that two-planet systems are all Hill-stable towards the end of simulations.   
On the other hand, for systems with more than two planets a dynamical instability may be triggered after the depletion of disk gas.  In that case, additional long-term dynamical evolution must be considered, which is beyond the scope of this work. We therefore restrict our sample to two-planet systems. 

Since Kepler planets are detected by transit and only their radii are measured, we use \cite{Weiss2014}'s mass-radius relationship to obtain the planet masses:
  \begin{equation}
\frac{M_p}{M_{\oplus}} =\begin{cases}
 {2.69  (R_p/R_{\oplus})^{0.93}}  & \mbox{ when $ R_p \geq 1.5 \ R_{\oplus} $ },  \\
 {   \rho_{\rm p}/ \rho_{\oplus}  (R_p/R_{\oplus})^{3} } & \mbox{ when $ R_p < 1.5 \ R_{\oplus}$ }, \\
 \end{cases}
 \label{eq:mass-radius}
\end{equation}
where $ \rho_{\rm p} = 2.43 + 3.39 (R_p/R_{\oplus})$. 
    
 The rapid disk depletion timescale  ($\tau_0$) in \eq{taudep} is numerically set to be $10^{4} \yr$. The first $10^{4} \yr$ simulation represents the gas-rich phase evolution  when  planets undergo inward migration and get trapped in a stable resonant state. As long as ${\dot M}_{\rm g0}$ remains fixed, $r_{\rm c}$ will stay the same and the planets will keep their orbits. When the disk starts to disperse ($t> \tau_0$), however, planets migrate outward with the cavity.   The simulation is terminated  when $t= \tau_0 + 10 \taud$ so that the gas density  is $10$ e-foldings of the initial value.

Each simulated system is assigned  one specific set of initial disk and stellar B-field parameters ($\mdot$, $\taud$, $B_{\star}$). Since the early disk and stellar properties are not well constrained, we obtain them by random sampling from a distribution. The distribution that we consider in this section is referred to as the fiducial distribution D.1 (\tb{diffrun}):  the disk dispersal timescale ($\taud$) is logarithmically sampled between  $10^{2} \yr $ to $10^{5} \yr$, the stellar B-field strength is linearly sampled  between $0.3$ to $3  \kG$, and the initial gas disk accretion rate is sampled logarithmically between $10^{-8} {\Msyr}$ to ${\dot M}_{\rm g, max}$ (see below).

Concerning the initial gas disk accretion rate ($\mdot$), the minimum value corresponds to  the typical value of young T Tauri stars, while the maximum value is a function of $B_\star$. 
Since the cavity radius ($r_{\rm c}$) is a decreasing function of the disk accretion rate (\eq{r_cavity}), 
 $r_{\rm c}$ would be very close to the host star when  $\mdot$ is very high.  The planets are at risk of being engulfed or tidally disrupted by their central stars. We therefore truncate the disk at $0.02 \AU$, twice the  radius of the young star ($2R_{\star} = 4R_\odot$), which, from \eq{r_cavity} corresponds to an accretion rate of   
\begin{equation}
  {\dot M}_{\rm g, max} = 2.8 \times 10^{-7} \left(\frac{B_{\star}}{1\kG} \right)^2  \Msyr. 
\end{equation}

The disk lifetime is a few Myr whereas the timescale of late stage disk dispersal is at least one order of magnitude shorter  \citep{Williams2011}.  Physically, the stellar UV and X-ray irradiation efficiently removes the disk gas by photo-evaporation, starting at around $1$ AU from the central star \citep{Alexander2006}. We focus on the  super-Earths in the inner disk where the gas is therefore drained quickly by viscous diffusion ($\sim$$10^4-10^{5}$ yr for $\alpha =10^{-2}$). In addition, disk winds could further reduce the depletion timescale \citep{Suzuki2010}.  For simplicity, we only mimic the depletion of the gas disk by considering a range of $\taud$, from $10^{2} \yr $ to $10^{5} \yr$.

\subsection{Period ratio distribution} 
 \begin{figure}[tb]
    \includegraphics[scale=0.64, angle=0]{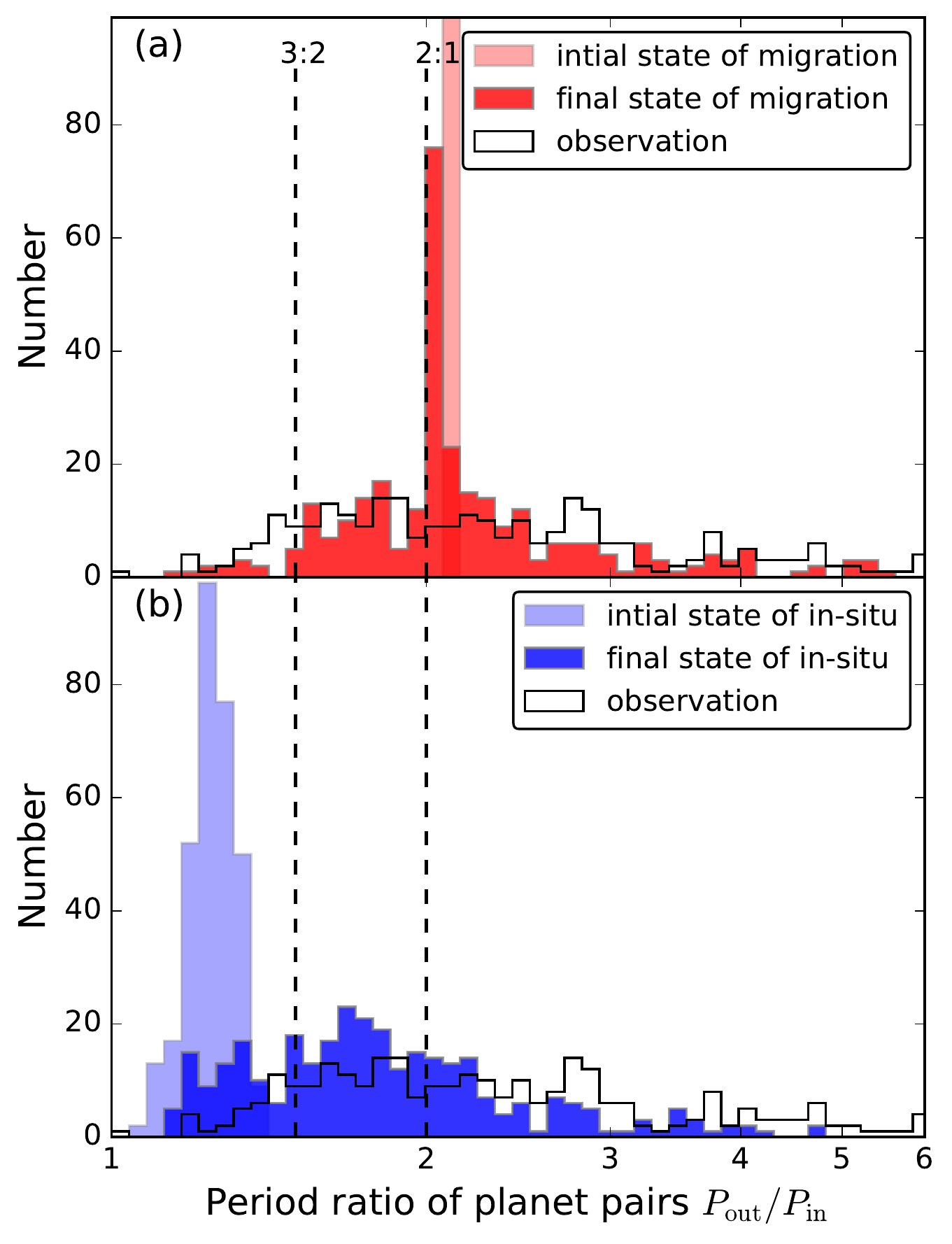}
    \caption{
        Histogram of the outer-to-inner planet period ratio. The upper and lower panel show the results from the migration (red) and the \textit{in-situ} (blue) scenario, respectively. In each scenario, both the initial (light) and final states (dark) of the simulations are given. The black line is the observation data for comparison.  Vertical lines indicate the $2$:$1$ and $3$:$2$ MMRs.} 
\label{fig:hist}
\end{figure}

In \fg{hist} we present the result of our simulations (fiducial case) in terms of the period ratio histogram. The upper and lower panel illustrate the migration and the \textit{in-situ} scenario, respectively. Three distributions are shown: the observed Kepler systems (black), the distribution before disk dispersal (light), and the final distribution after disk dispersal (dark).

In \fg{hist}a, the planets are assumed to have formed through the migration scenario. Initially, they start at a period ratio $2.1$ (\se{scenarios}). After  disk dispersal we find that nearly one third of the planets remain close to the $2$:$1$ resonance. 
However, the rest of the planets undergo substantial  migration.  The rebound  plays a crucial role during disk dispersal, which moves planets outward and increases their period ratios. Finally planets end up at period ratios over a wide range compared to their initial values.

In \fg{hist}b planets are assumed to have formed through the \textit{in-situ} scenario. Initially, they are separated by six to ten mutual Hill radii (approximately corresponding to period ratios between $1.1$ to $1.4$, see \se{scenarios}).  The rebound also increases the period ratios of planets. We find that more planets end up inside  the $2$:$1$ resonance for the \textit{in-situ} scenario than the migration scenario. Also, no strong resonance concentration is seen in the  \textit{in-situ} scenario.

 \subsection{Parameter dependence}

%\begin{figure}[H]
\begin{figure*}[tb]
\centering
  \includegraphics[scale=0.33, angle=0]{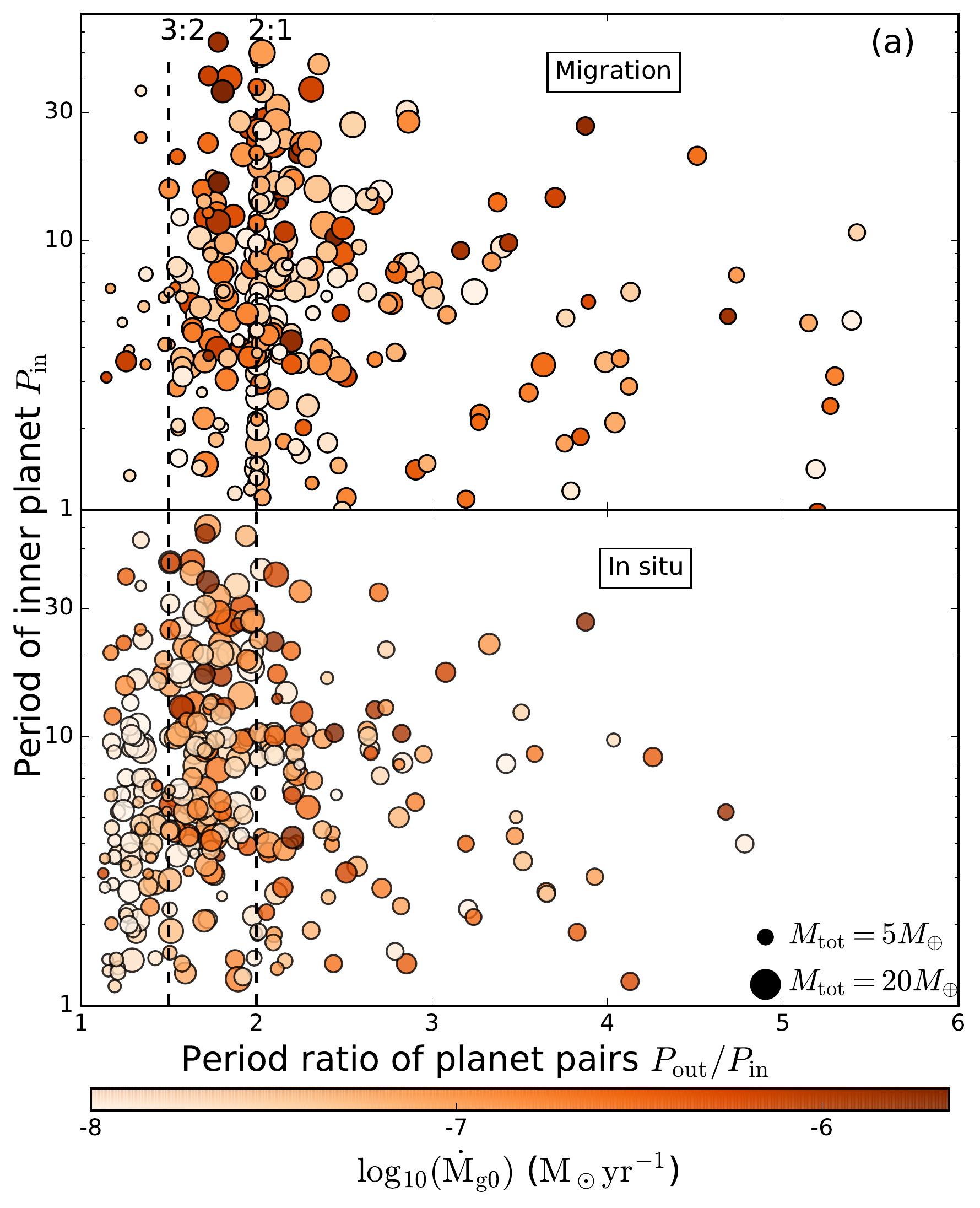}
    \includegraphics[scale=0.33, angle=0]{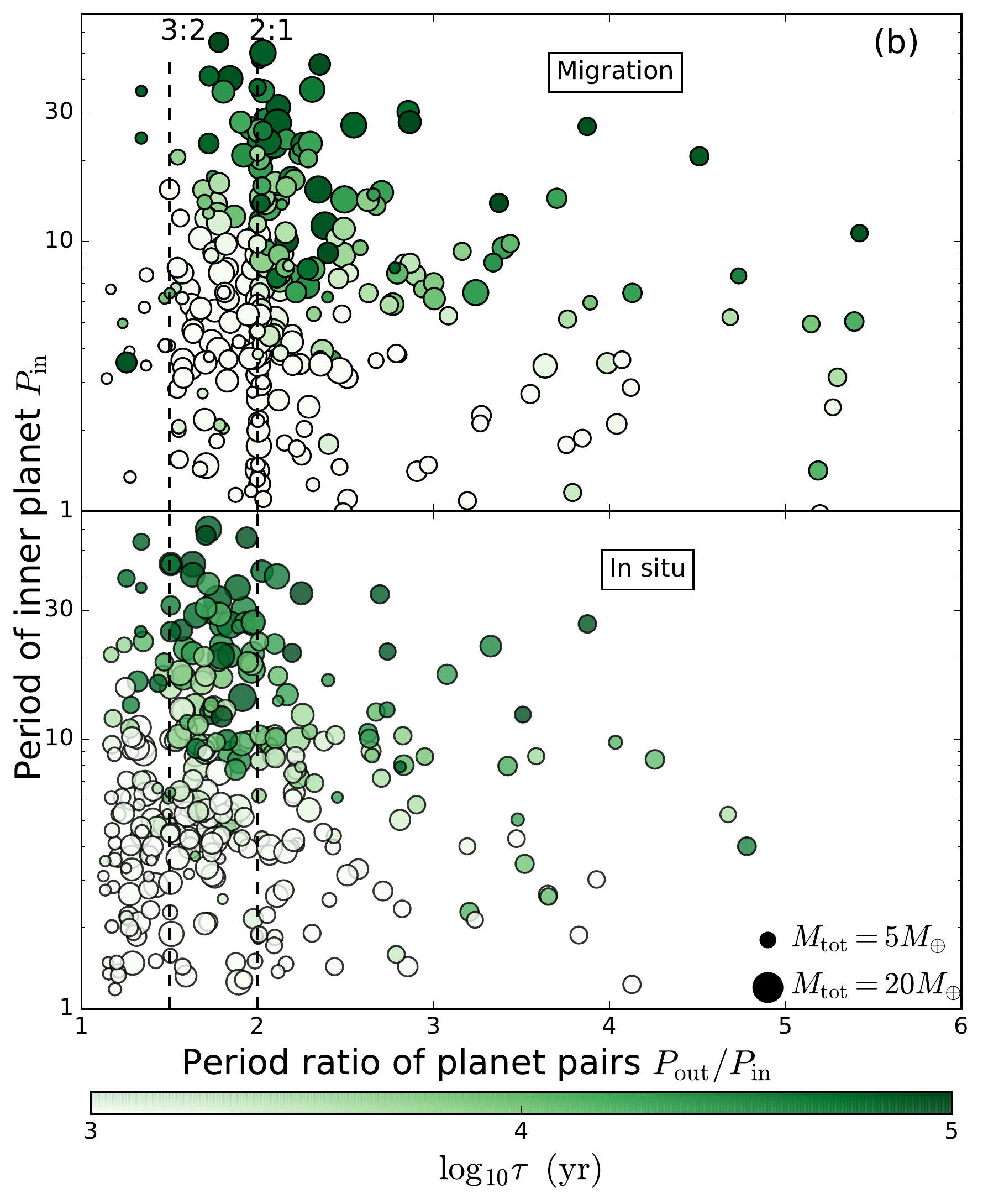}
      \includegraphics[scale=0.33, angle=0]{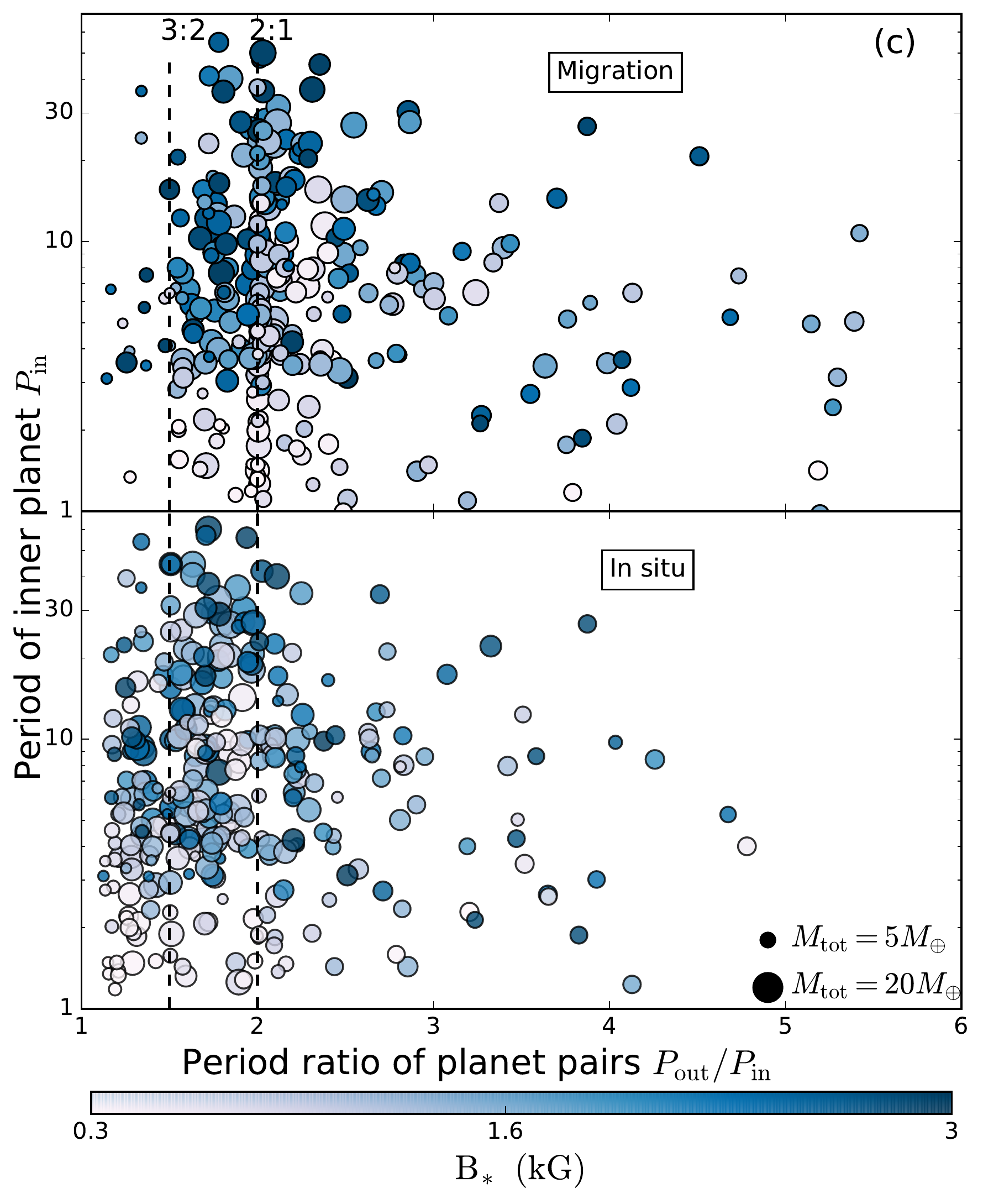}
    \caption{
    Scatter plot of the inner planet period  and the outer-to-inner planet period ratio. Each dot represents the final state of one particular planetary system. 
    The color of the dot corresponds to $\dot M_{\rm g0}$ (orange in panel a), $\taud$ (green in panel b), or $B_{\star}$ (blue in panel c), respectively. The  size of the dot represents the total mass of two planets.  Vertical lines mark the $2$:$1$ and $3$:$2$ MMRs.  }
\label{fig:para}
\end{figure*}

\subsubsection{Initial gas disk accretion} 
\label{sec:mdot}

In \fg{para} the results of the simulations are shown as a scatter plot of the inner planet period and the outer-to-inner planet period ratio. Three panels illustrate the dependence of these results on different parameters by colorbars.

In \fg{para}a, the color corresponds to the value of the initial gas disk accretion rate.  We find that for the migration scenario, systems with  high accretion rates are more frequently found  inside  the $2$:$1$ resonance than those with low accretion rates.  A large number of planets in low accretion disks (white) are clustered at period ratios $\gtrapprox 2$,  because they are trapped in $2$:$1$ resonances in the early gas-rich phase. They also migrate modestly during disk dispersal and end up at period ratios around $2$. On the other hand, planets in high accretion disks ($\gtrsim2-3\times 10^{-7} \Msyr$) are more likely to bypass the $2$:$1$ resonance to end up in, for example,  the $3$:$2$ resonance.
In addition, resonance trapping depends on planet mass.  We find that systems with low-mass planets ($M_{\rm p } \simeq M_{\oplus}$) are able to bypass the $2$:$1$ resonance even in low accretion disks, which is consistent with the analysis and simulations of \cite{Ogihara2013}.

In the \textit{in-situ} scenario planets are initially  in a compact configuration (all planets are within the 3:2 MMR). In this case, we find that planets end up at a smaller period ratio in low accretion disks (white) than those in high accretion disks (orange). This is because the rebound cannot efficiently move planets outward and increases their period ratios in low accretion disks. 

The variation of the disk accretion rate has two consequences on the inner planet period. On the one hand, $r_{\rm c}$ truncates the disk further out when the disk accretion rate is lower, so the planet stops its inward migration at a larger distance from the star. On the other hand, planets cannot migrate outwards strongly due to weak rebound in low accretion disks. As a result,  we do not find a clear dependence of the inner planet period on the disk accretion rate.   

In conclusion, we find that for both scenarios planets in disks with a low accretion rate tend to stay in their original orbits.  The period ratios show a wider distribution with increasing disk accretion rate, whereas the period distribution does not show a clear trend on the disk accretion rate.

\subsubsection{Disk depletion timescale} 
\label{sec:taud}
In  \fg{para}b we plot the same simulation result, but this time the color corresponds to the disk depletion timescale $\tau_d$, which ranges from $10^2$ to $10^{5} \yr$. We find that the planets with a long depletion time (green)  generally end up at longer periods and larger period ratios than those with a short depletion time (white).  This behavior is simply due to the rapid depletion of gas, which limits the ability of the planet to migrate outwards. On the other hand, when the disk depletes slowly, both planets are able to migrate outwards, in sync with the cavity expansion, for a substantial time. The planet stops the outward migration when its migration timescale becomes longer than the depletion timescale (Sect. 3 of paper I).  As a result, systems with a short depletion timescale only rebound modestly while systems with a long depletion timescale can significantly expand their orbits.

\subsubsection{Stellar B-field strength} 
In \fg{para}c we also examine the stellar B-field strength dependence  on the final configuration of planetary systems. In general, we find that systems with large $B_{\star}$ (blue) end up at longer orbital periods.   A larger stellar B-field truncates the disk at a larger cavity radius $r_{\rm c}$ (\eq{r_cavity}). Therefore, the final period of the inner planet ($P_\mathrm{in}$) becomes longer in systems with a stronger stellar magnetic field.  However, the stellar magnetic field does not affect  the period ratio of planets.

To summarize,  the disk dispersal time affects both the inner planet period and the period ratio. In  longer depletion disks planets migrate further out and end up at larger period ratios.  The stellar magnetic field strength mainly determines the orbital period of the  planet.  Systems with stronger magnetic field strengths tend to have larger orbital periods.  Planets in disks with low accretion rates tend to preserve their orbital configurations. Their period ratios can increase over a wide range as the disk accretion rate increases. These trends are independent of the planet formation scenario (initial condition).

 %\section{Comparison between statistical surveys and the observations}
\section{Comparison to Kepler two-planet statistics}
\label{sec:optimization}

In this section, we focus on the comparison between simulations and observations. The observed data is selected from our Kepler sample, limited to two-planet systems (\se{setup}).  Since the initial disk and stellar B-field parameters are poorly known, we have performed simulations for both the fiducial case (\se{compare_fiducial}) and other underlying  parameters' distributions (\se{compare_diff}).  Comparing the simulations to the observed periods and period ratios, we can infer which distribution and which scenario best fit the observations.
\subsection{ Fiducial case}
\label{sec:compare_fiducial}

\subsubsection{Period ratio distribution} 
In \fg{hist} we have also plotted the observed period ratio of the Kepler sample (black open histograms). It can be seen that even though the initial period ratio distributions for the  two scenarios are very different, magnetospheric rebound significantly reduces this disparity and results in two final period ratio distributions that are overall  certainly similar to the observations.The clear exception is the $2$:$1$ resonance excess in the migration scenario. We do not obtain a good match to the observations in that case.  An additional mechanism to move the planet out of resonances is needed (see the discussion of \se{compare_diff}).  

\subsubsection{Planet mass ratio} 
\label{sec:ratio}  
In \fg{obs},  both simulation (upper) and observation (lower) are shown as a scatter plot of the inner planet period and the outer-to-inner planet period ratio.  The color corresponds to the outer-to-inner planet mass ratio. A red (blue) dot indicates that the outer planet is more (less) massive  than the inner planet. 

The simulations ( upper panel of \fg{obs}) show that the planets' period ratio correlates with their mass ratio. In both scenarios, systems with larger outer-to-inner planet mass ratios (red) show a wide spread in period ratios. Conversely, systems with smaller outer-to-inner planet mass ratios (blue) have smaller period ratios.   Specifically, in the migration scenario planets show a pile-up around the $2$:$1$ MMR and a small fraction of  the systems end up at period ratios smaller than $2$. However, in the \textit{in-situ} scenario, the dominant period ratio is smaller than $2$.

Although the planets get trapped into resonance in the gas-rich phase, magnetospheric rebound rearranges the dynamical configuration of planets during disk dispersal.   In the case of $M_{\rm out}/M_{\rm in}$$>$$1$,  the more massive outer planet is able to migrate outwards and decouple  resonance when the inner planet falls into the cavity.  As a result, the rebound increases the planets' period ratio. However, in the case of $M_{\rm out}/M_{\rm in}$$<$$1$,  once the inner planet enters the cavity, the less massive outer planet cannot migrate outwards as fast as the expanding cavity. Thus, it also falls into the cavity and ends up at nearly resonant orbit with the inner planet.   Therefore, with the rebound mechanism, the planets' period ratio  correlates with the mass ratio of the planets.

\begin{figure}[t]
    \centering
    \includegraphics[scale=0.5, angle=0]{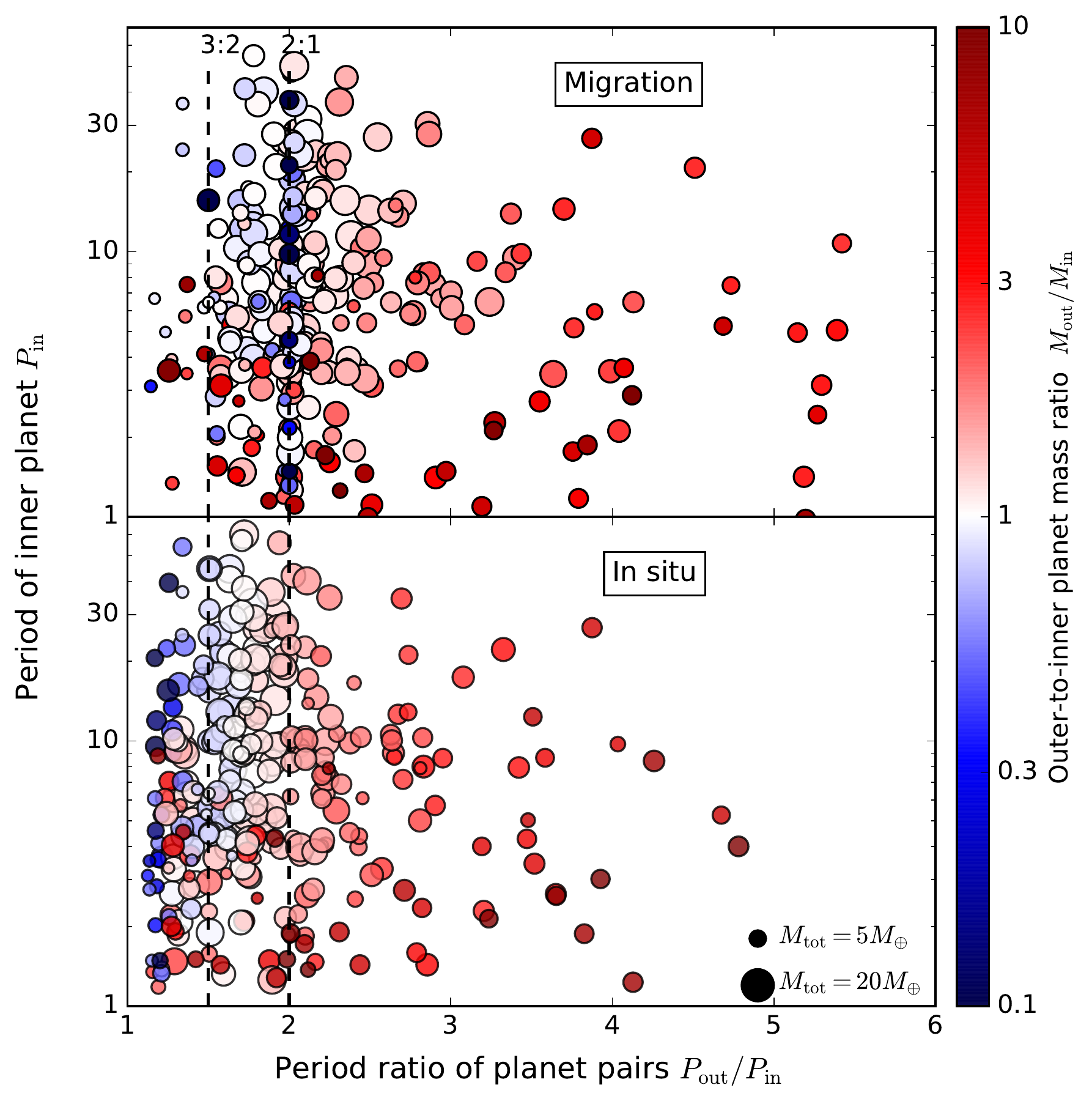}
    \includegraphics[scale=0.46, angle=0]{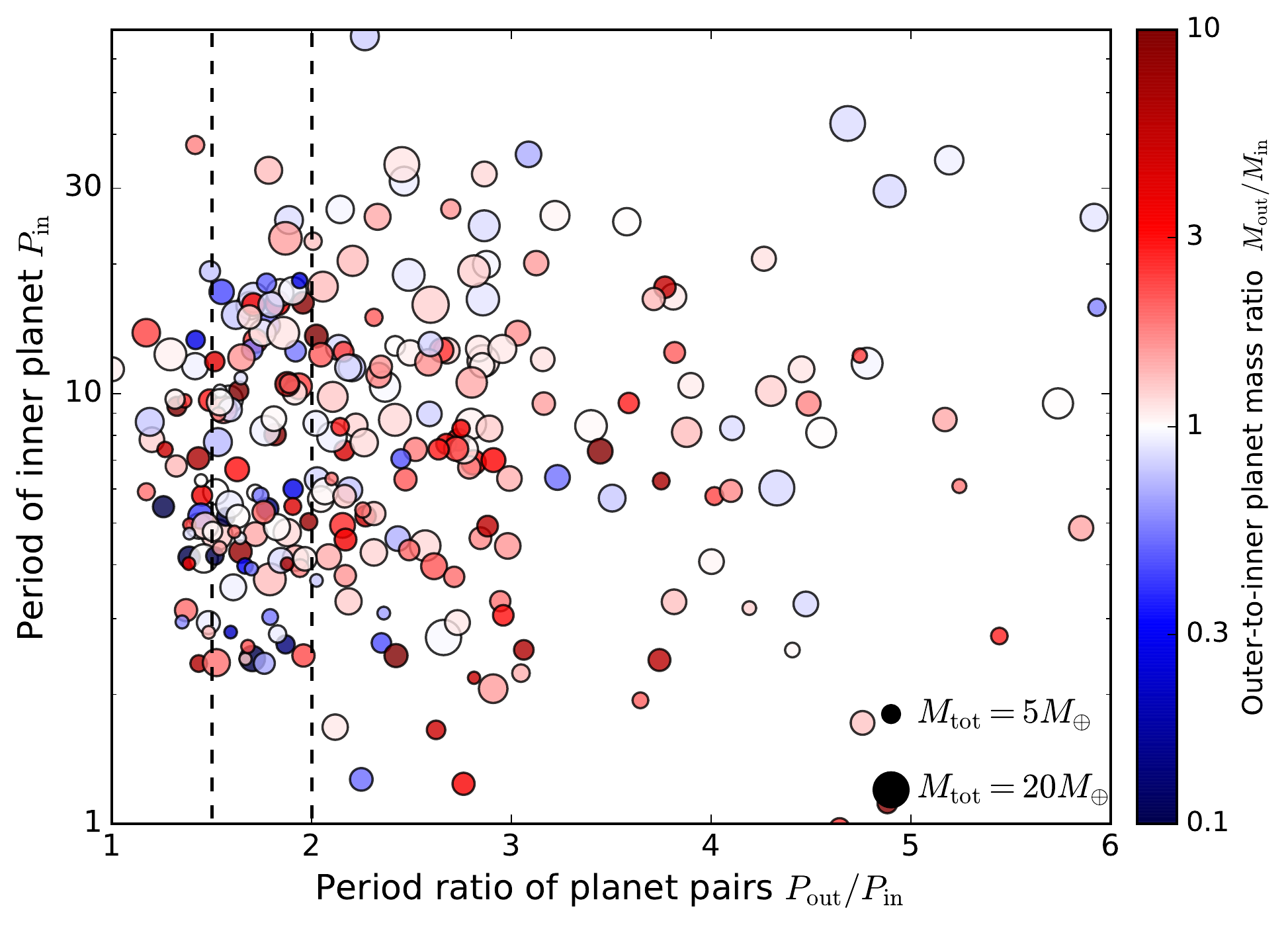}
    \caption{ Scatter plot of the inner planet period and the outer-to-inner period ratio from simulations (upper panel) and  from our Kepler sample (lower panel).
 The size of the dot indicates the total mass of planets in an individual system, and the color (blue to red) corresponds to their outer-to-inner mass ratio (small to large).  The masses of planets are calculated from their radii by \cite{Weiss2014}. 
    }
    \label{fig:obs}
\end{figure} 

Comparing simulations and observations (\fg{obs}), the period ratio-mass ratio correlation shown by the simulations is not reflected in the observations, which, on first sight, do not exhibit any correlation between period ratio and mass ratio.\footnote{In order to determine whether such correlation exist in the data, a proper bias correction needs to be conducted \citep{Steffen2015,Christiansen2015}, which is beyond the scope of this work.}

We propose several explanations for this discrepancy.  First, both the observed and simulated distributions would be affected by any scatter and uncertainty that enters the conversion of \textit{Kepler} radii into masses (\eq{mass-radius}). It is, however, likely that the true mass will deviate from the estimated mass based on \eq{mass-radius}. For instance,  \cite{Wolfgang2016} propose a probability model for the mass-radius relation rather than precise one-to-one mapping. Conceivably, many of the white dots ($M_{\rm in} \simeq M_{\rm out}$) shown in \fg{obs} could in fact be red.  A better mass estimation or independent mass determination is needed to further verify our prediction. Second, some of  the observed two-planet systems may have had more embryos at an early stage. After disk dispersal, planet scattering and mergers could produce a system with two planets or multiple planets with only the inner two close enough to be detectable. Systems that have experienced such a post-dispersal instability could have their period ratio--mass ratio correlation erased, due to the chaotic nature of the scattering.  Finally, the magnetospheric rebound mechanism may not have operated because requirements have not been fulfilled (Paper I). For example, the well-ionized inner disk condition may not be true for some Kepler systems during their early evolution stage. Unfortunately, it is hard to presently assess which \textit{Kepler} systems were most conducive for the rebound effect and those that were not.

\subsection{Effect of different input distributions}
\label{sec:compare_diff}

In \se{compare_fiducial} we have only explored one particular distribution (D1 in \tb{diffrun}).  
Here we study two additional  distributions (D2-3 in \tb{diffrun}), where  D2 and D3 represent new simulations with different distributions for $\dot{M}_{\rm g0}$ and $\tau_{\rm d}$, respectively.  The `low disk accretion rate' runs (D2) feature a reduction of the minimum disk accretion rate  by a factor of $10$ (($\dot M_{\rm g0,min}= 10^{-9} \Msyr$) compared to the fiducial runs (D1) .   
Similarly, D3 and D1 have an identical distribution for the disk accretion rate, but the distribution for the disk depletion timescale is narrower and skewed towards large $\tau_d$ in D3, which ranges from  $10^4$ to $10^{5} \yr$.

\begin{table*}
    \centering
    \caption{Statistical survey}
    \begin{tabular}{lllllllllllll|l|}
        \hline
        \hline
       Distribution ID & Description  & $\dot M_{\rm g0}  $ ($\Msyr$)   & $\tau_{\rm d}$ ($ \yr$)    & $B_{\ast}$  ($\kG$)\\
        \hline
      D1 & fiducial   & [$10^{-8}$, $\dot M_{\rm g0,max}$]  & [$10^2$, $10^{5}$] & [$0.3$, $3.0$]   \\
      D2    & low disk accretion rate      &   [$10^{-9}$, $\dot M_{\rm g0,max}$]& [$10^2$, $10^{5}$]  & [$0.3$, $3.0$]   \\
      D3  &  long disk decay time &    [$10^{-8}$, $\dot M_{\rm g0,max}$]  & [$10^4$, $10^{5}$] & [$0.3$, $3.0$]   \\
    \hline
        \hline
    \end{tabular}
    \label{tab:diffrun}
\end{table*}

\begin{figure*}[htb]
\includegraphics[scale=0.51]{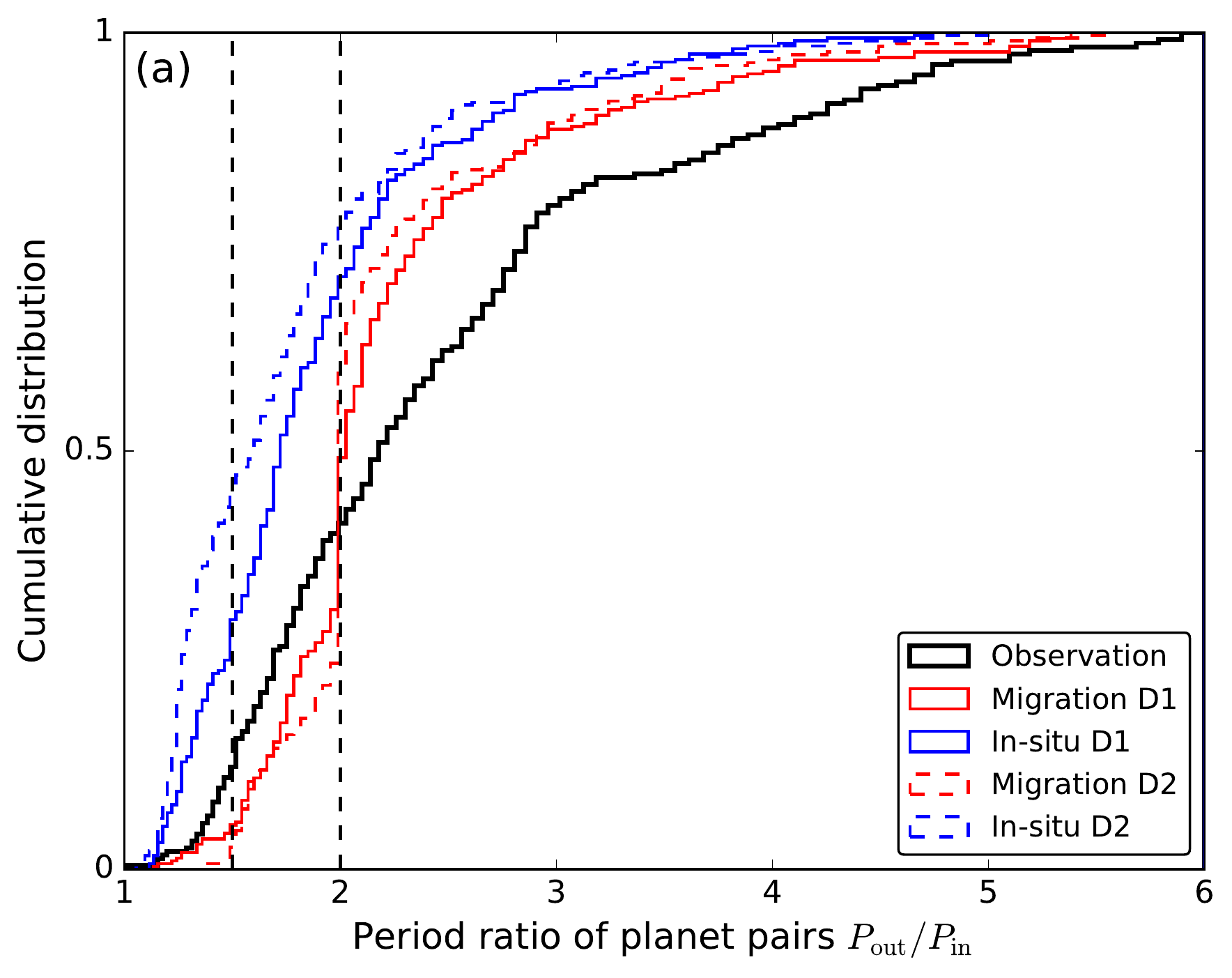}
\includegraphics[scale=0.51]{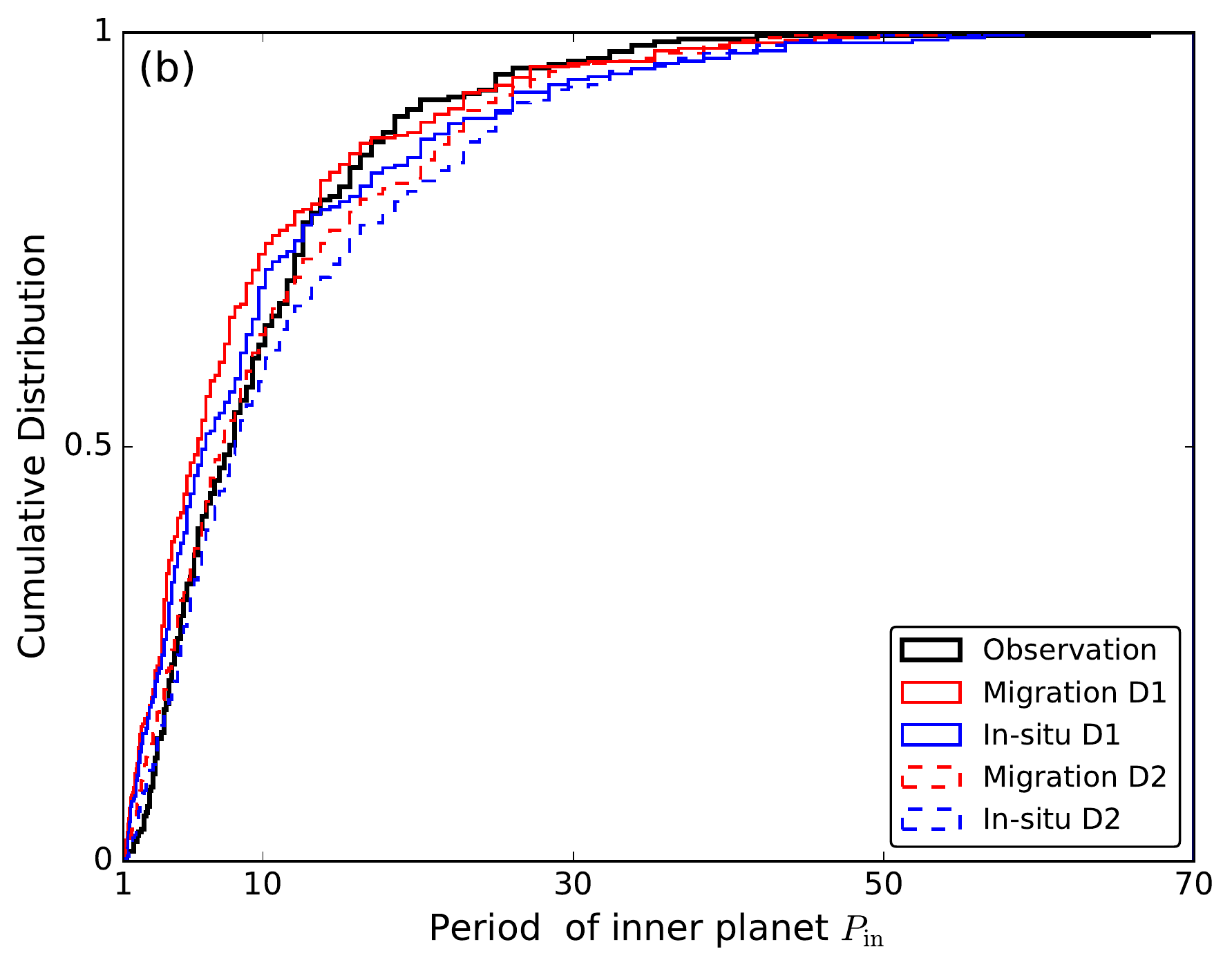}
\includegraphics[scale=0.51]{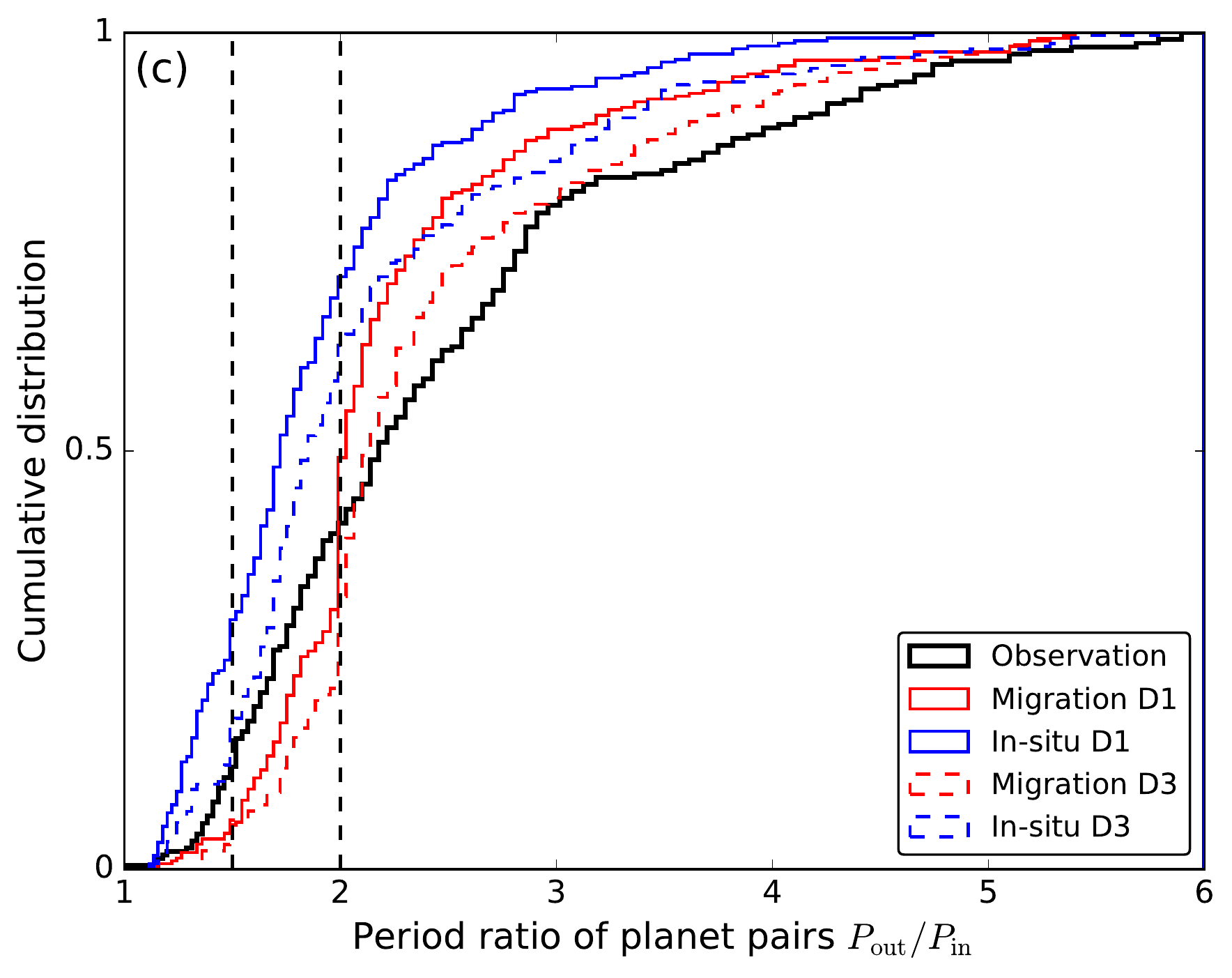}
\includegraphics[scale=0.51]{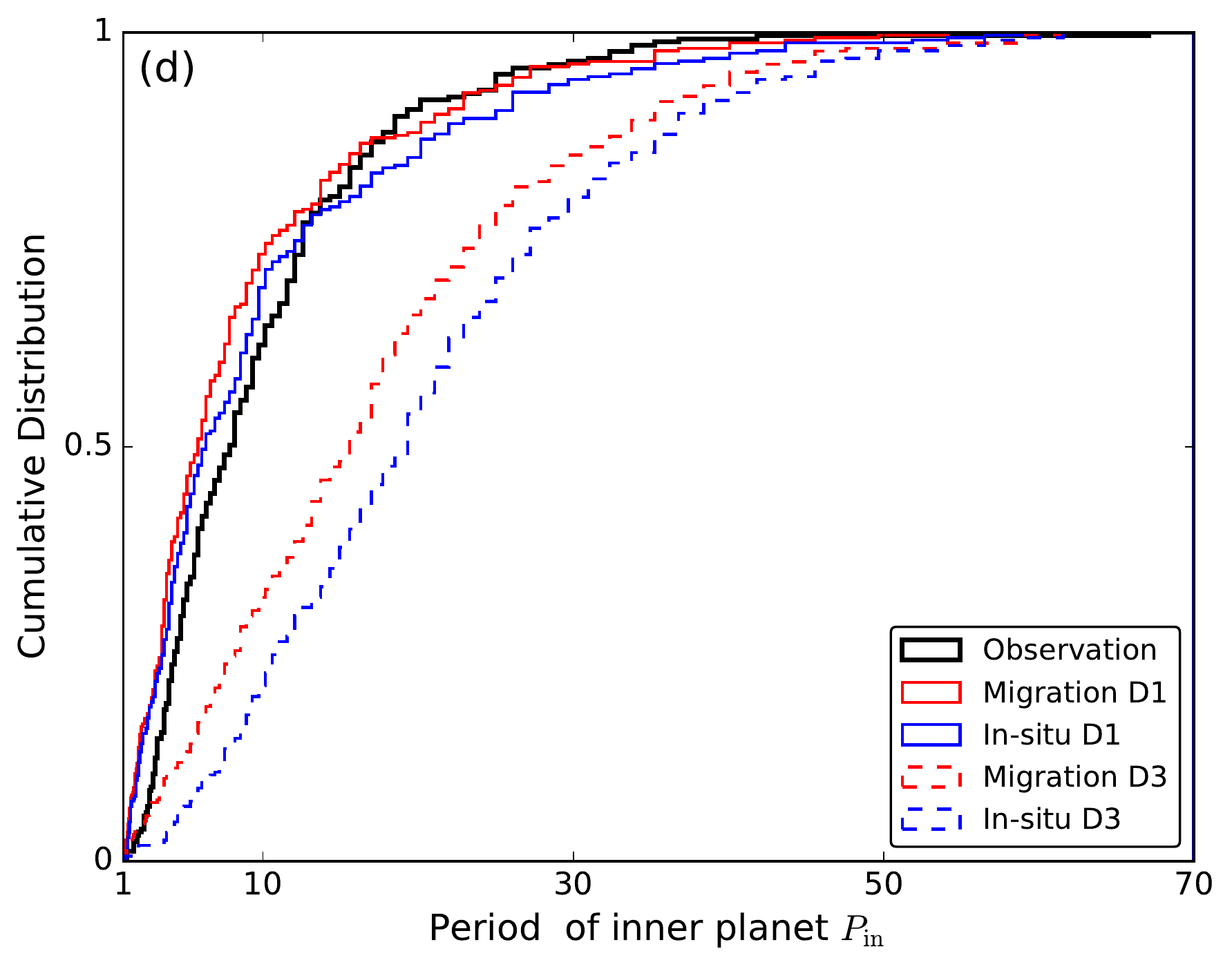}
\caption{
Cumulative distribution of planet period ratio (left) and inner planet period (right). The black solid line represents the observations, whereas the red and blue lines show the results from the migration and the \textit{in-situ} scenarios, respectively. The solid  lines are the result of D1 in all four panels. Dashed lines give the results of D2 in panel (a) and (b), and D3 in panel (c) and (d), respectively. Vertical lines mark the $3$:$2$ and $2$:$1$ MMRs. 
}
\label{fig:distribution}
\end{figure*}

The results are shown in \fg{distribution} as cumulative distributions for the inner planet period and period ratios, respectively.  
For the migration scenario (red curves), we find that less planets end up at period ratios smaller than $2$ in the low accretion distribution runs (D2) than in the fiducial case (D1) (\fg{distribution}a). This is because planets in low accretion systems cannot easily bypass the $2$:$1$ resonance (\se{mdot}). Hence, a lower disk accretion rate results in even more planets ending up at the $2$:$1$ resonance, deteriorating the fit to the observations. 
For the \textit{in-situ} (blue curves) scenario, more planets  stay at short period ratios  in D2, which also deteriorates the match to the observations.  Because the magnetospheric rebound  becomes weaker in a lower accretion disk,  the period ratio of planets thus cannot increase much.

In \fg{distribution}b, we find that planets in the low accretion runs (D2) have an inner planet period distribution that is similar to the fiducial runs (D1).    The cavity radius ($r_{\rm c}$) becomes larger when $\dot{M}_{\rm g0}$ is lower. However, rebound also becomes weaker when  $\dot{M}_{\rm g0}$ is lower. Combining these two effects, we find that  the inner planet period distribution is almost unaffected  by $\dot{M}_{\rm g0,min}$ and that both runs fit the observed $P_\mathrm{inn}$ well.

Given our finding that a lower $\dot{M}_\mathrm{g0,min}$ results in a worse fit to the period ratio distribution, the reader may wonder whether a larger  $\dot{M}_{\rm g0, min}$ would  improve the fit  to the observations.   We adopt  the minimum value of the disk accretion rate in the fiducial runs, $\dot{M}_{\rm g0,min}=10^{-8} \Msyr$, for two reasons. First, it is the typical observed value among solar-mass, T Tauri stars.  We do not expect systems to start their rapid disks dispersal at a very large $\dot{M}_{\rm g0}$. Second,  $r_{c}$ is closer in  when  $\dot{M}_{\rm g0}$ is larger (\eq{r_cavity}).  A larger $\dot{M}_{\rm g0, min}$  results in the cavity radius close to the stellar truncation radius ($0.02 \AU$), and the disk accretion rates are all close to the maximum value. We find this also does not improve the match to the observations.

Finally, we consider the comparison between  the narrow $\tau_{\rm d}$ distribution runs (D3) and the fiducial case (D1). The time $\tau_{\rm d}$ represents how fast the disk gas depletes. A larger $\tau_{\rm d}$ means that the disk depletes more slowly and that more gas is left to drive the planet migration and  `fuel' the rebound.  In \fg{distribution}c and \fg{distribution}d, we find that both the inner planet period and the planet period ratio are generally larger in the runs where the disk depletion timescales are longer (D3). As discussed in \se{taud}, rebound can move planets outwards when their migration timescale is shorter than the disk depletion timescale. Therefore,  planets end up with larger period ratios in disks where the disk gas depletes more slowly.     For the period ratio distribution (\fg{distribution}c), the simulations with the narrow $\tau_d$ distribution (D3) slightly improve the fit to the observations, both for the migration and the \textit{in situ} scenario. However, in Fig. 4d we see that more planets in D3 end up at larger periods, which is inconsistent with the observations.

Optimizing the parameter distributions  to obtain the best match to the observations is not the aim of this work.  Instead, we have mainly focused on the general trends seen in the simulations and their implications in light of the magnetospheric rebound model.  For the investigated parameter distributions, we find that neither scenario can perfectly explain the observational features.  
The main shortcoming of the  \textit{in-situ} scenario is that  planets do not end up at very large period ratios (\fg{hist}b).   
It is important to note, however, that  we only consider two-planet systems. More embryos that form early could undergo giant impacts in the gas-free phase after the termination of the magnetospheric rebound.    This chaotic process can reduce the number of planets by collisions, and increase their final separations \citep{Hansen2012,Hansen2013,Pu2015}. Widely-separated planet pairs may form in this way.
On the other hand, the problem with the migration scenario is that it produces an over-abundant population of planets at the $2$:$1$ resonance.
Additional mechanisms, such as stellar tides \citep{Lithwick2012,Delisle2014b}, stochastic disk torques \citep{Rein2012,Batygin2017}, or dynamical instabilities  \citep{Izidoro2017} are necessary to move planets out of resonances.

\section{Conclusions}
\label{sec:conclusion}

In this paper, we have investigated the magnetospheric rebound model that we introduced in Paper I in light of the  Kepler data.
We have studied the relation between the observed Kepler systems  and modern planet formation theory from a statistical perspective. Two formation scenarios have been investigated -- migration and  \textit{in-situ} -- which result in different orbital properties in the early, gas-rich disk phase.
We  simulate the evolution of the Kepler systems during the gas disk dispersal for these two scenarios. The influence of each model parameter ($\dot M_{\rm g0}$, $\tau_{\rm d}$, and $B_{\star}$) on the inner planet period and period ratio distributions has been statistically investigated (\se{parameter}). We have explored different sets of initial disk and stellar B-field distributions. The comparison between simulations and  the observations is conducted in \se{optimization}.
 
The major findings of this paper are summarized as follows: 
\begin{enumerate}[1.]
     \item Planets in disks with low accretion rates do not much change their orbits after the disk dispersal.  Period ratios can increase significantly as disk accretion rates increase (\fg{para}a).
     \item  The disk dispersal time affects both the planet period and the period ratio. Planets in disks with  longer depletion timescales end up with larger orbital periods and period ratios (\fg{para}b).  
     \item The stellar magnetic field strength mainly determines the period of the planet. Systems with  stronger stellar magnetic fields tend to have longer orbital periods (\fg{para}c). 
     \item The magnetic rebound model predicts a correlation between the period ratios of planets and their mass ratios: systems with more massive outer planets end up at large period ratios whereas those with lighter outer planets preferentially stay at small period ratios. However, present-day Kepler systems do not seem to exhibit such correlation (\fg{obs}).      
     \item   The migration scenario overproduces $2$:$1$ resonance planets. However, planets are less  dominated by resonances compared to the gas-rich disk phase. In the \textit{in-situ} scenario, planets increase their orbital periods and end up at larger period ratios. But, compared to the observations, there is still a lack of planets with very large period ratios (\fg{hist}).     
     \item  Magnetospheric rebound significantly diminishes the erstwhile differences between the two formation models. Therefore, it is hard to conclude which formation model best fits the Kepler two super-Earth planet population (\fg{hist} and \fg{distribution}).

 \end{enumerate}

 Overall, the magnetospheric rebound model  provides an explanation for  a mass ratio-period ratio dependence, for
   the significant decrease in the number of resonant planets, and provides a vehicle for planets at large orbital periods and period ratios. These model predictions qualitatively agree with the observations. Nevertheless, we also give explanations why our model does not always match  the observations (\se{compare_diff}). As mentioned, the model outcome  depends on the masses of the planets and particularly on the mass ratio. The masses of most Kepler planets, however, are difficult to infer by follow-up radial velocity surveys since the host stars are too faint. Future space missions, such as Transiting Exoplanet Survey Satellite (TESS) and Planetary Transits and Oscillations of stars (PLATO), have the ability to detect planets down to Earth-size around  nearby bright stars. For these bright planets, follow-up RV surveys will be able to measure the masses accurately. Consequently, TESS and PLATO will provide larger and more precise data sets to verify the trends predicted by the magnetospheric rebound model.

\begin{acknowledgements}
We thank  Douglas N.C. Lin, Carsten Dominik, Allona Vazan, Jiwei, Xie and Gijs Mulders for useful discussions and the referee for valuable comments.
 B.L.\ and C.W.O\ are supported by the Netherlands Organization for Scientific Research (NWO; VIDI project 639.042.422).  

\end{acknowledgements}

\bibliographystyle{aa}
\bibliography{reference}

\end{document}